\documentclass[prb,twocolumn]{revtex4}
\usepackage{epsfig}
\usepackage{graphicx}
\usepackage{amsfonts,amssymb}
\usepackage{amsmath}
\usepackage{color}
\usepackage {times}
\definecolor{refcolor}{rgb}{1.0,0.0,0.0}

\newcommand{\be}{\begin{equation}}
\newcommand{\ee}{\end{equation}}   
\newcommand{\bea}{\begin{eqnarray}}
\newcommand{\eea}{\end{eqnarray}}
\newcommand{\ba}{\begin{array}}
\newcommand{\ea}{\end{array}}

\renewcommand{\j}{{\bf j}}

\newcommand{\q}{{\bf q}}
\renewcommand{\k}{{\bf k}}

\begin{document}

\title{Fermi arcs and pseudogap phase in a minimal microscopic 
model of $d$-wave  superconductivity}

\author{Dheeraj Kumar Singh$^{1}$, Samrat Kadge$^{2}$, 
Yunkyu Bang$^{3,4}$, and Pinaki Majumdar$^{2}$} 

\affiliation{$^1$School of Physics and Materials Science, 
Thapar Institute of Engineering 
and Technology, Patiala-147004, Punjab, India}

\affiliation{$^2$Harish-Chandra Research Institute, HBNI,
Chhatnag Road, Jhunsi, Allahabad 211019, India}

\affiliation{$^3$Department of Physics, POSTECH, Pohang, 
Gyeongbuk 790-784, Korea}

\affiliation{$^4$Asia Pacific Center for Theoretical Physics, 
Pohang, Gyeongbuk 
790-784, Korea} 

\date{\today}

\begin{abstract}
We show conclusively that a pseudogap state can arise at $T > T_c$, 
for reasonable pairing interaction strength, from order parameter 
fluctuations in a  two dimensional minimal model of $d$-wave 
superconductivity. The occurrence of the pseudogap requires neither 
strong correlation nor the presence of competing order.  We study a 
model with attractive nearest neighbor interaction and establish our 
result using a combination of cluster based Monte Carlo for the order 
parameter field and a twisted-boundary scheme to compute the 
momentum-resolved  spectral function. Apart from a dip in the density
of states that characterizes the pseudogap, the momentum and frequency 
resolution on our effective lattice size $\sim 160 \times 160$ allows 
two major conclusions: (i)~at $T < T_c$, despite the presence of 
thermal phase fluctuations the superconductor has only nodal Fermi 
points while all non nodal points on the normal state Fermi surface 
show a two peak spectral function with a dip at $\omega =0$, and
(ii)~for $T > T_c$ the Fermi points develops into arcs, characterized
by a single quasiparticle peak, and the arcs connect up to recover 
the normal state Fermi surface at a temperature $T^* > T_c$. We show 
the variation of $T_c$ and $T^*$ with coupling strength and provide 
detailed spectral results at a coupling where $T^* \sim 1.5T_c$.
\end{abstract}

\maketitle

\section{Introduction}

Experiments on the underdoped cuprates were the first to focus 
attention on a `pseudogap' phase~\cite{robinson}.  It was observed 
that a gap persists in the quasiparticle excitation spectrum 
above the superconducting transition temperature, $T_c$.  
Unlike the conventional superconductors, there exists another 
temperature scale, $T^* > T_c$, for the disappearance 
of the antinodal gap \cite{ding, loeser, renner, norman}. 
An intriguing aspect of the spectral properties in the pseudogap 
(PG) phase is the appearance of Fermi arcs above $T_c$, while the 
$T < T_c$ phase shows only nodal Fermi points despite thermal phase 
fluctuations.  The  $T_c < T < T^*$ window shows partial gapping of 
the Fermi surface. The length of the Fermi arcs increases with 
temperature until the normal state Fermi surface (FS) is recovered
\cite{yoshida, kanigel1,kanigel2,hasimoto,kondo,sato,doiron} at $T^*$.

Theoretical work on this problem has explored multiple possibilities. 
One focuses on the fluctuation in amplitude of the $d$-wave 
superconducting (dwSC) order parameter as the origin of PG behaviour
~\cite{randeria,emery,norman1,franz,kwon,atkinson}. 
Another - the `semi-classical approximation' (SCA) - 
attributes 
the Fermi-arc formation to a square-root singularity along 
the FS in the spectral function that results 
from the Doppler shift of the 
quasiparticle energy in the presence
of supercurrent in the PG phase~\cite{franz,berg}.
Later study, however, indicated that
such singularity may be an artifact of the SCA, which itself 
may be unjustifiable because the quasiparticles are 
massless near the nodes in the PG region~\cite{khodas}. 
A recent work~\cite{sumilan} considers electron self-energy 
correction due 
to the exchange of a Cooper-pair fluctuation at finite 
temperature. A crucial input in the theory, the 
correlation length $\xi(T)$, 
was assumed to have a phenomenological 
Berezinskii-Kosterlitz-Thouless 
(BKT)~\cite{olsson} form while the temperature dependent 
dwSC order parameter $\Delta(T)$ was 
adopted from the ARPES measurements. 

Despite the exploration of various physical mechanisms, 
the microscopic origin of the PG phase in the high-$T_c$ cuprates 
remains highly debated. An exact diagonalization (ED) + 
Monte-Carlo (MC) based study in a microscopic model~\cite{mayr} 
pointed out that the PG phase may not exist for a realistic value 
of interaction, $V \sim t$ (where $t$ is the hopping scale) 
in a minimal microscopic model. On the other hand, another work 
using a relatively larger system size suggested otherwise~\cite{zhong}. 
However, both  approaches suffered from a system size that was not
adequate to examine 
the momentum-resolved spectral function and establish the 
existence of Fermi arcs in the $T > T_c$ window. Another 
approach stressed on competing  dwSC and antiferromagnetic 
order~\cite{paramekanti} using MC simulation of 
a Landau-Ginzburg (LG) functional and also 
considered the presence of quenched 
disorder~\cite{alvarez}. The order parameters obtained were
used in a microscopic model to show the existence of the 
PG phase. The topological aspect of this transition has also
been discussed~\cite{varma}. Finally, recent work indicates 
that the onset of the PG phase may be accompanied by the 
appearance of nematic order~\cite{sato,stajic,murayama,
mukhopadhyay,kivelson,choubey,tu}.

In this richly varied field it may be useful to first
establish conclusively what spectral features can emerge
from purely $d$-wave pairing fluctuations, and only then
build in additional effects. In this spirit we explore the 
impact of classical thermal
fluctuations of the $d$-wave order parameter field on
the electronic spectrum. In these studies the ED+MC method is the 
most frequently used approach.  However, the use of the approach is 
limited by the relatively small lattice size that is accessible 
owing to the high computational cost~\cite{mayr,zhong}. This 
is a serious hindrance for examining the spectral properties,
especially the momentum-resolved spectral
functions along the normal state Fermi surface, an exercise
essential to examine the pseudogap phase.

In this paper, we adopt an approach
which uses a combination of a cluster based MC method
\cite{kumar} to access thermal fluctuations on reasonably large
sizes, and a
twisted-boundary condition (TBC) scheme \cite{salafranca} 
to obtain high resolution spectra.  
For a small system size, with poor momentum resolution,
it is difficult to determine the extent of Fermi arc. 
This is because only a small fraction of momentum points 
fall on the FS or are close to it.
Therefore, we first obtain equilibration for size
$L_{MC} \times L_{MC}$ ($L_{MC} = 20$) 
and for spectral calculations 
employ TBC - repeating a equilibrated configuration 
$L_{tw}$ = 8 times along both $x$ and $y$ 
direction. Then, the momentum-resolved spectral 
function is obtained by using Bloch's theorem for 
a lattice of effective size $L_{eff} = L_{MC} \times L_{tw}$,
{\it i.e}, $160 \times 160$.
We summarise our main results below in terms of the single
particle spectral function $A({\bf k}, \omega)$.

(i)~We show the existence of a PG phase above $T_c$ in a minimal 
model for dwSC without invoking the presence of any competing 
order, suggesting that the key difference between the PG phase 
and pure dwSC may be the absence of phase correlation in the 
former.  (ii)~For $T < T_c$ there is a two peak structure in 
$A({\bf k}, \omega)$ for all momenta on the normal state Fermi 
surface, except the nodal points - where a quasiparticle peak is
visible. The spectral weight
$A({\bf k},0)$ on this contour falls sharply away from the nodal 
points. (iii)~For $T > T_c$ the single peak quasiparticle feature
is visible over a larger part of the normal state Fermi 
surface, forming  `Fermi arcs' around the nodal points.
The Fermi arcs increase in length and connect up to create
the normal Fermi surface at $T = T^*$, where the two peak
feature at the antinodal point also collapses.

\section{Model and method}

\subsection{Model}

We consider a minimal two dimensional 
electron  model 
\be
H = -\sum_{ {\bf i} \j  \sigma} t_{{\bf i} \j}
d^{\dagger}_{{\bf i}\sigma} d_{{\bf j}\sigma}
- \mu \sum_{{\bf i} } n_{{\bf i}} -|V|
\sum_{\langle{\bf i}{\bf j}\rangle} n_{\bf i} n_{\bf j}
\ee
The first term is the kinetic energy which
includes both first ($t$) and second ($t^{\prime}$)
neighbor hoppings. We set $t=1$. We choose 
$t^{\prime}$ = -0.4 so as to reproduce the experimentally
observed Fermi surface~\cite{damascelli}.
In the second term, $\mu$
is the chemical potential, which is chosen to correspond 
to the band filling $n \sim 0.9$. Finally, the last term
describes the nearest-neighbor attractive interaction 
responsible for dwSC pairing. The interaction parameter 
$V \sim 1.0$ is chosen so that it is consistent with the 
nearest-neighbor antiferromagnetic coupling $J \approx 
4t^2/U \sim 1$. We have fixed $V = 1.2$ for all the
 calculations unless
stated otherwise.

\subsection{Monte Carlo strategy}

The effective Hamiltonian, below,  
employed in the simulation process can 
be obtained formally via a Hubbard-Stratonovich transformation
of the intersite interaction in the $d$-wave pairing channel
and assuming the pairing field 
$\Delta^{\delta}_{\bf i}$ to be `static', {\it i.e}, classical.
This is equivalent, structurally, to a `mean field' like 
decoupling 
of the interaction, without any additional assumption about
homogeneity and phase correlation among the $\Delta^{\delta}_{\bf i}$.
The $\Delta^{\delta}_{\bf i}$ are allowed both amplitude and
phase fluctuations.
We have ignored other possible decouplings, for example related
to charge density wave, etc.
\begin{eqnarray}
H_{eff} &=& -\sum_{ {\bf i}, \delta^{\prime} , \sigma} 
t_{{\bf i},{\bf i} + \delta^{\prime}} 
d^{\dagger}_{{\bf i}\sigma} d_{{\bf i}+\delta^{\prime} \sigma}
- \mu \sum_{{\bf i} } n_{{\bf i}}  \cr
&& - \sum_{{\bf i},\delta} 
[(d^{\dagger}_{{\bf i} \uparrow}d^{\dagger}_{{\bf i}
+\delta \downarrow} + 
d^{\dagger}_{{\bf i}+\delta \uparrow}d^{\dagger}_
{{\bf i} \downarrow})\Delta^{\delta}_{\bf i} + h.c.] + H_{cl}
\cr
H_{cl} &=& ~~\frac{1}{V}\sum_{\bf i}|\Delta^{\delta}_{\bf i}|^2 
\end{eqnarray}
$\delta^{\prime}$ refers 
to both the first and second nearest neighbors 
whereas $\delta$ to only the first neighbor. 
The superconducting gap function defined on the link 
is a complex classical field and can be expressed as
$\Delta^{\delta}_{\bf i} = |\Delta_{\bf i}|
e^{i \phi^{\delta}}$. For simplification, $|\Delta_{\bf i}|$ is
treated as a site variable while 
$\phi^{\delta}$ $({\delta} = x, y)$ as a link variable. 

The equilibrium configurations $\{\Delta_i, \phi^x_i, \phi^y_i \}$ 
are determined using the 
Metropolis algorithm, which involves
updating of the configuration according to the
distribution 
\be 
P \{ \Delta_i, 
\phi^x_i, \phi^y_i \} \propto Tr_{dd^{\dagger}}e^{-{\beta H_{eff}}}.
\ee 
For an update at a given site 
the Hamiltonian corresponding to a cluster of size $L_C \times L_C$ around the update site 
is diagonalized (using periodic boundary condition),
instead
of the full  20$\times$20 system Hamiltonian. We use
$L_C=6$.  This approximation is 
based on the assumption that the effect of the proposed 
change at the update site decreases quickly with distance 
as one moves away~\cite{kohn}. Benchmarked earlier for 
spin-fermion and other similar models, it leads to a significant 
reduction in update cost from $\sim N^3$ to $\sim N^3_c$
for a system and cluster with $N$ and $N_c$ sites,
respectively~\cite{kumar}. After equilibriation we consider 
a superlattice constructed using dwSC fields configurations 
on the $L_{MC} \times L_{MC}$ system and study the 
spectral properties by using Bloch's theorem as 
discussed in later subsections.    

For the interaction parameter considered in this
work, we start MC simulations at a temperature $T \sim 0.05t$ 
way above the dwSC transition temperature and reduce the
temperature to cool down the system in steps of $\Delta
T = 0.0013t$. The small temperature step ensures that 
the system avoids any metastable states during the annealing
process.

\subsection{$T_c$ determination}

From the equilibrium configurations at a given temperature we can
calculate the long-range phase correlation $\Phi(L_{MC}/2,0)$:
\be
\Phi(s_x,s_y) = \frac{1}{N}\sum_{\bf i}
 \langle e^{i \phi^x_{\bf i}} 
e^{i \phi^x_{{\bf i}+{\bf s}}}\rangle. 
\ee
The $T_c$ for dwSC is determined 
from the  the rise of $\Phi(L_{MC}/2,0)$ 
on reducing $T$ from high temperature.
To keep track of any 
deviation from the dwSC state, we calculated another useful 
correlation function 
$
\Psi = \frac{1}{N}\sum_{\bf i} \langle 
e^{i \phi^x_{\bf i}} e^{i \phi^y_{{\bf i}}}\rangle.
$
A negative $\Psi $ indicates whether the state obtained in the 
equilibration is a dwSC state or not. For the
 interaction parameters considered in the current work, only 
 dwSC state is obtained.

Earlier proposal for the determination of the onset
temperature $T^*$ of PG phase focused on the short-range phase 
correlation function such as $\Phi(1, 0)$,
with $\Phi(1, 0) \sim 0.1$ set as a criterion for $T_c$.
The inference~\cite{mayr} was that for a realistic $V \sim 1$ 
no PG is possible.  However, the 
validity of such assumptions was not checked by examining
 the spectral
function, which was challenging owing to the finite size effect. 

\subsection{Spectral features}

Once the thermal equilibrium is achieved, we use twisted 
boundary condition to calculate the density of states. For instance, 
$t_{{\bf i} {\bf j}}
 \rightarrow t_{{\bf i} {\bf j}} e^{- i (q_x a_x + q_y a_y)}$, where 
$q_x,q_y = 0, 2\pi/N_l, 4\pi/N_l,...2\pi(N_l-1)/N_l$. $N_l = L_{tw} = 8$ 
is the number of lattices in the superlattice,
\textit{i.e}, number of repetition along $x$ and 
$y$ direction of the lattice under consideration in 
both the directions. Note that we set $a_x = a_y = 1$. Similarly,
$\Delta^{\delta}_{\bf i} = |\Delta_{\bf i}|
e^{i \phi^{\delta}} = |\Delta_{\bf i}|
e^{i \phi^{{\bf i} \j}} = |\Delta_{\bf i}|
e^{i \phi^{{\bf i} \j}}e^{-i (q_x a_x + q_y a_y)} $ at the boundaries. 
Then, the density of states (DOS) is calculated as  
\bea
N(\omega) &=& \sum_{\q, \lambda, {\bf i}} 
(|u_{\q,\lambda}({\bf i})|^2\delta
(\omega - E_{\q , \lambda}) \nonumber\\
&& ~~+ |v_{\q,\lambda}({\bf i})|^2\delta(\omega + E_{\q, \lambda})) 
\eea
where $E_{\q, \lambda}$ are the eigenvalues of the Hamiltonian
 obtained from Eq.1 using 
Bogoliubov-Valatin transformation. $|u_{\q,\lambda}\rangle$
and $|v_{\q,\lambda}\rangle$ form the 
eigenvectors of the Hamiltonian. 
The single particle spectral function is calculated as 
\bea
A(\k, \omega) & = & \sum_{\q, \lambda} ( |\langle \k  
|u_{\q,\lambda}\rangle |^2 \delta(\omega - E_{\q, \lambda}) \nonumber\\ 
&&~~+  |\langle \k  |v_{\q,\lambda}\rangle |^2 \delta(\omega +
 E_{\q, \lambda})) 
\eea
where
\be
\langle \k|u_{\q,\alpha}\rangle = \sum_l \sum_i \langle \k |
l, i \rangle \langle l, i | u_{\q,\lambda} \rangle.
\ee
$l$ is the superlattice index and $i$ is a site index 
within the superlattice. 

\begin{figure}[b]
\vspace{-.5cm}
\begin{center}
\includegraphics[width=8.7cm,height=4.1cm,angle=0]{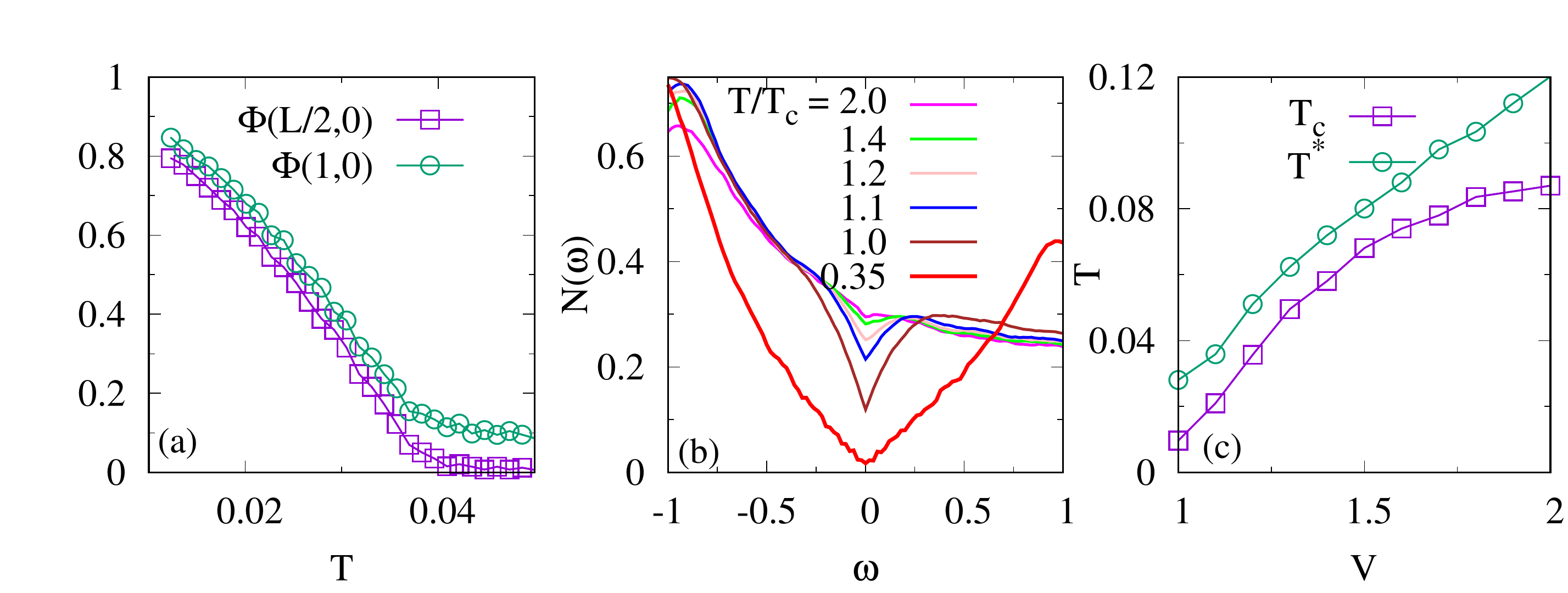}
\end{center}
\vspace{-.7cm}
\caption{(a)~Long-range and short-range phase correlation function
as a function of temperature. 
(b)~Evolution of DOS as a function of temperature.
The dip in the $V$-shaped DOS continues to exist beyond $T_c$
and can be noticed up to $T \sim 1.5T_c$.
(c)~$T$ and $T^*$ as
a function of interaction $V$, where the critical interaction
strength $V_c \sim$ 1.0 for the onset of $d$-wave superconductivity.
} 
\label{struct}
\end{figure}
\begin{figure}[t]
\begin{center}
\hspace{-9mm}
\includegraphics[width=8.9cm,height=4.5cm,angle=0]{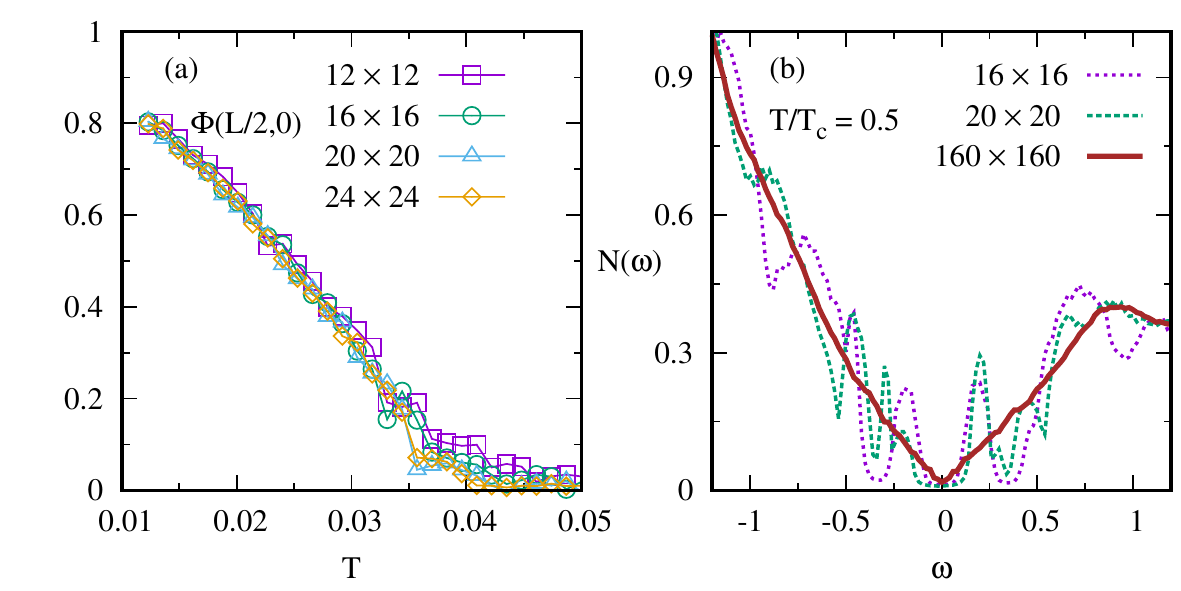}
\vspace{-5mm}
\end{center}
\caption{(a) Phase correlation functions as a function of temperature
for different lattice sizes $L_{MC} \times L_{MC}$ with $L_{MC} =$
12, 16, 20 and 24. (b) DOS for $L_{MC}$ = 16 and 20  with $L_{tw}$
= 1. For comparison, DOS is shown also for $L_{MC} = 20$ and
$L_{tw} = 8$, i.e, an effective lattice size of 160 $\times$ 160.}
\label{struct1}
\end{figure}

\section{Results}

Fig.\ref{struct}(a) shows both long  and short-range phase 
correlations, $\Phi(L_{MC}/2,0)$ and $\Phi(1,0)$, as a function 
of temperature. On decreasing temperature, $\Phi(L_{MC}/2,0)$ 
starts to rise from a nearly zero value at $T_c \sim 0.035t$. 
$\Phi(1, 0)$ is $ \sim 0.1$ for a large temperature window above
 $T_c$ for the model in this work.

One of the earliest  experimental
signatures of the pseudogap was a dip in the 
density of state (DOS) persistent
above $T_c$, obtained in the STS measurement~\cite{renner}. 
We find a qualitatively similar dip as shown in Fig.\ref{struct}(b), 
which retains it’s $V$-shape structure even above $T_c$. The
dip becomes shallow enough to become unnoticeable only at
temperature $T \sim 1.5T_c$, for $V=1.2$. 
We estimate $T^*$ as the temperature at which
the dip in the DOS becomes unnoticeable. 

Based on that 
criterion, our estimate of $T^*$ reveals it’s dependence 
on $V$ to be similar to that of $T_c(V)$ in the
intermediate coupling regime, with the $T^*(V)$ curve running
almost parallel to the $T_c$ curve (Fig.\ref{struct}(c).
Importantly, $T^*/T_c$ grows with a decrease 
in the interaction parameter $V$, {\it i.e}, 
the relative temperature 
window for pseudogap phase is larger~\cite{micnas} for 
realistic $V \sim 1$.

Fig.\ref{struct1} shows the dependence of the long range 
phase correlation
function and density of states on the lattice size. The phase
correlation shows only a relatively small suppression with
increasing lattice size in the vicinity of the onset temperature 
of dwSC (Fig.\ref{struct1} (a)). 
For temperature $T = 0.5T_c$,
Fig.\ref{struct1}(b), shows the lattice size dependence of the
density of states at temperature $T = 0.5T_c$. We compare
sizes 16 $\times$ 16, 20 $\times$ 20, and results for an
an effective lattice size 160 $\times$ 160 (using twisted 
boundary conditions). The 
effective lattice is equivalent to an equilibrated field 
configurations on a 20 $\times$ 20 lattice size repeated
8 times along both along $x$ and $y$ directions. The 
finite size artifacts are absent in the latter while they
are significant in the two smaller sizes.

\begin{figure}[b]
\vspace{-5mm}
\begin{center}
\psfig{figure=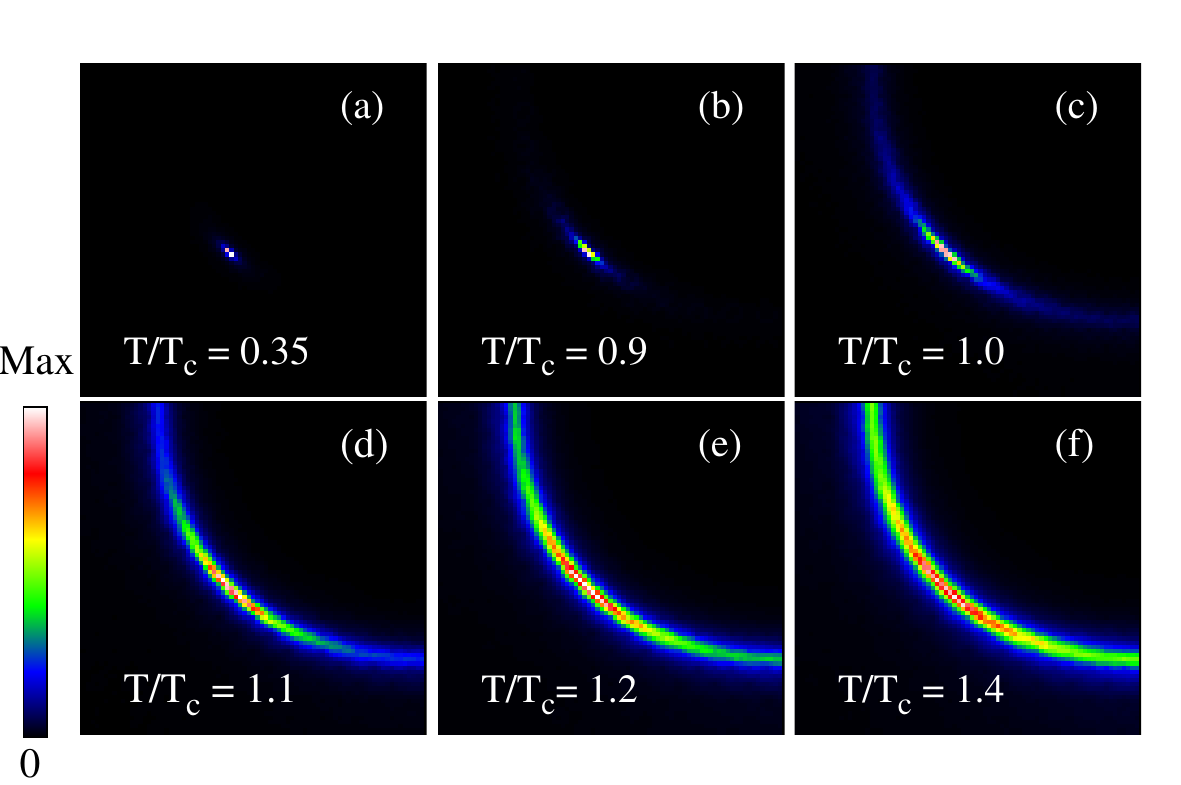,width=9.0cm,height=6.0cm,angle=0}
\end{center}
\vspace{-5mm}
\caption{(a)-(f) Evolution of quasiparticle spectral weight 
$\mathcal{A}({\k}, \omega)$ for $\omega = 0$ as a function of temperature.
$k_x$ and $k_y$ are along horizontal and vertical directions, respectively,
with each having range [0, $\pi$].  $\mathcal{A}({\k}, 0)$ increases
continuously with temperature away from the nodes. However, it is only
tiny fraction
($\sim$ 1$\%$) of that in the vicinity of node, a feature noticeable even
beyond $T_c$, which is an indication for a preformed Cooper-pair state
existing up to a very high temperature.}
\label{qpe}
\end{figure}

Fig.\ref{qpe} shows that well below 
$T_c$ the spectral weight is concentrated at the
nodal point. On approaching $T_c$, the spectral 
weight builds up continuously at points near the nodal points 
along the normal state FS. We will examine the 
energy dependence
of the spectral functions further on.
The weight at points on the nominal FS 
away from the nodal point remains only a tiny fraction ($\sim$ 
1$\%$) of that at the node. The process of spectral 
weight build up continues beyond $T_c$.  The spectral weight 
remains highest near the nodal points, 
and smallest near the  antinodal, even beyond $T^{*}$, which 
clearly indicates the existence of Cooper-pairs without any
phase coherence between them. Above $T_c$,
the pseudogap can be expected to fill quickly first near 
the nodal points, and then away from them, until the 
whole of the normal state Fermi surface appears.
 
\begin{figure}[t]
\vspace{0mm}
\centerline{
\psfig{figure=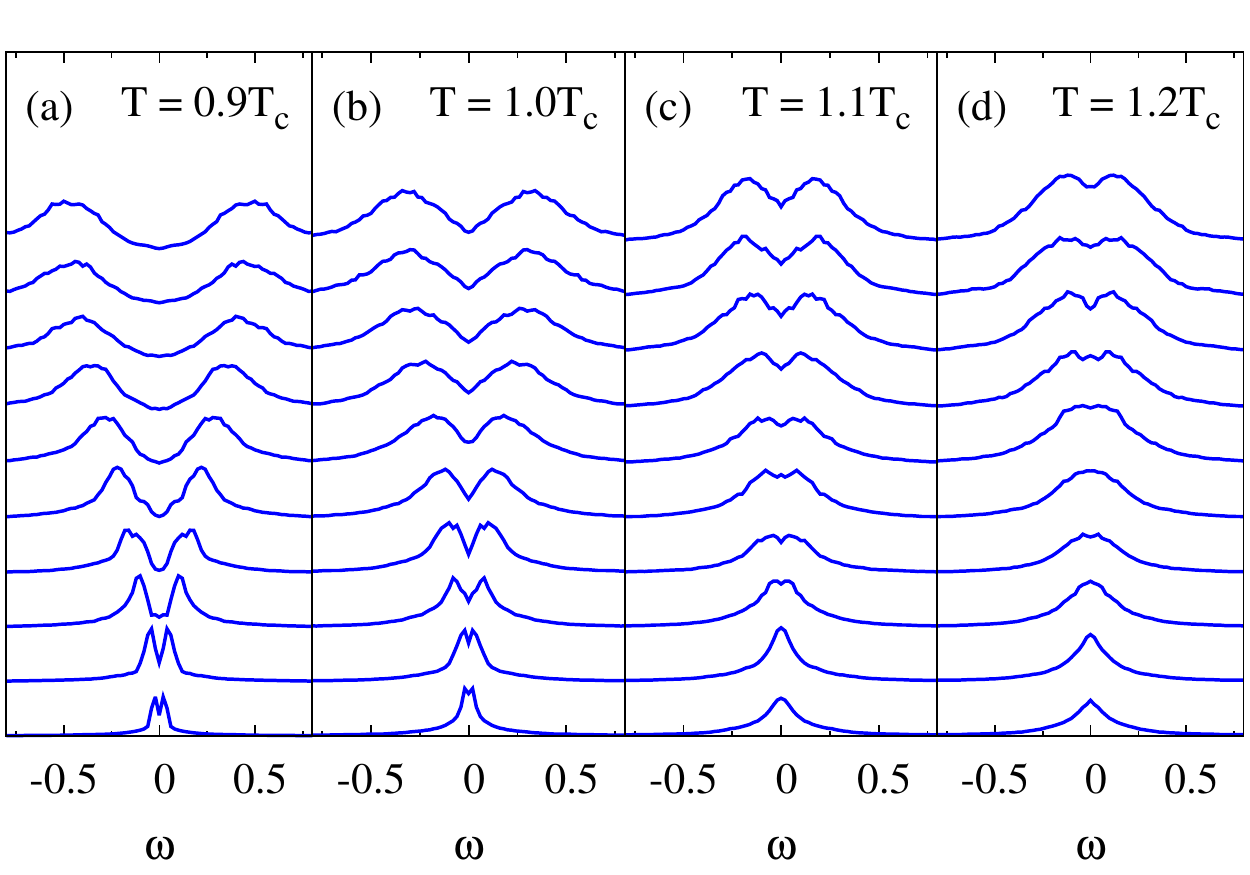,width=8.7cm,height=6.0cm,angle=0}}
\caption{Momentum resolved spectral function $\mathcal{A}({\k},\omega)$ 
at different temperatures calculated for several values of momentum $\k$
along the Fermi surface ($V $= 1.2). The bottom curve in each case 
corresponds to a point very close to the nodal point while the top curves 
to a point in the vicinity of antinode.  Nodal quasiparticles are 
protected against the phase fluctuation below $T_c$ and the Fermi arcs 
are formed above $T_c$ that continue to exist for $T \gtrsim 1.4T_c$. 
Asymmetry present in the spectral function, visible more when the gap 
is small, is an artifact of the finite size of the system. The peak 
height is scaled arbitrarily to enhance the visibility.}
\label{qpe1}
\end{figure}

To understand how the gaps are filled away from the nodal 
points either below or above $T_c$, we examine the 
momentum-resolved 
spectral function as plotted in Fig.\ref{qpe1} for the
 points ($\k$) along the normal state FS. Because of the 
finite size effect, most of the points are slightly away 
from the normal state
FS (within the range of $\Delta k_x$ = $\Delta k_y \lesssim 0.03$). 
 It introduces small asymmetry which is removed by plotting the 
symmetrized spectral functions $(\mathcal{A}({\bf k}, \omega) +
 \mathcal{A}({\bf k}, -\omega))/2$ instead.  

Several points are to be noted. First of all, the SC order parameter 
retains it's $d$-wave character below $T_c$, despite phase 
fluctuations.
There exists two-peak structure with a small dip at $\omega = 0$ 
for any non-nodal point, however close they are to the nodes, all 
the way up to $T_c$. It indicates that the Fermi points
 remain intact against the thermal phase fluctuations below $T_c$. 
 This is in contrast to the mean field 
picture in which the dwSC order parameter approaches
 zero all along the normal state FS while preserving it's $d$-wave 
 symmetry. Secondly, the spectral weight increases continuously,  
as the temperature increases at the points nearby to nodes, 
along the normal state FS. This is 
also accompanied with the disappearance of the dip associated 
with thermally broadened two-peak structure  and appearance 
of a single broad peak, starting from the points in the vicinity
 of nodal points and extending up to those near the antinodal 
 points, as $T$ increases beyond $T_c$ and reaches $T^*$. 
 Thus, the region $T_c < T < T^*$ is marked by the spectral 
 features which are in qualitative agreement with the ARPES measurements~\cite{kanigel2}.

We have found that the peak in the  dwSC amplitude 
distribution shows 
only a small shift across the entire temperature range 
considered. We find $\Delta_{an} (T_c) \approx 0.5 \Delta_{an}
(0)$, while $\Delta_{an} (T)$ does not decrease noticeably within
 the range $T_c \lesssim T \lesssim T^*$. The feature, 
which is indicative of the PG as a preformed Cooper-pair state 
without absence of any phase coherence, also agrees 
with the ARPES
 measurements according to which $\Delta_{an}(T)$ 
 remains independent of temperature for the entire range
 $0 < T < T^*$. For us, however, 
 $\Delta_{an}(T)$ shows a drop in size by nearly one half 
on approaching $T_c$ from below. 

\begin{figure}[t]
\centerline{
\psfig{figure=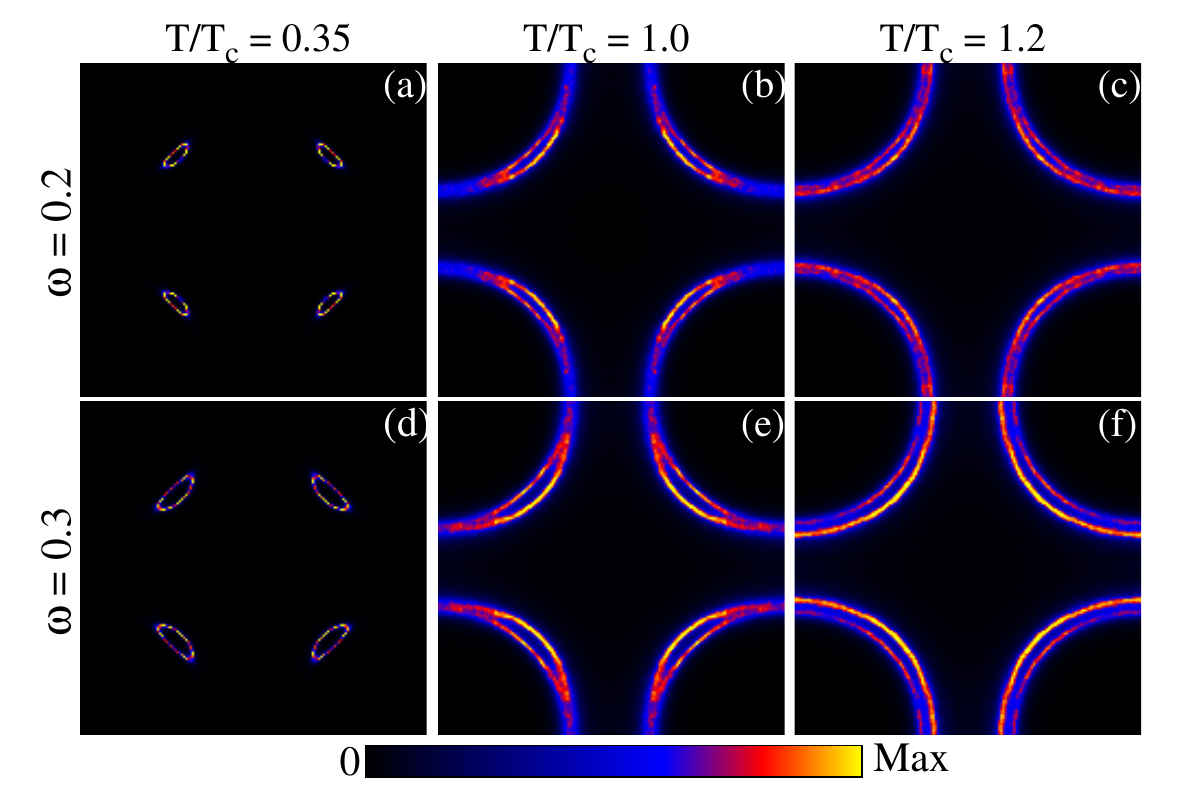,width=9.0cm,height=6.0cm,angle=0}}
\caption{Thermally averaged $\mathcal{A}(\k, -\omega)$+
$\mathcal{A}(\k, \omega)$ for $k_x$ and $k_y$ in the range 
[-$\pi$, $\pi$]. 
The columns are for temperatures $T/T_c = 0.35, 1$ and $1.2$, 
while the rows show data at $\omega = 0.2$ and $0.3$, 
respectively.} 
\label{qpe2}
\end{figure}

Fig.\ref{qpe2} shows the thermally-averaged spectral function 
$\mathcal{A}(\k, -\omega)$+ $\mathcal{A}(\k, \omega)$. The 
existence of banana-shaped constant energy surfaces can be 
seen nearly up to $T_c$. These  banana-shaped constant 
energy surfaces within the octet model have been used to
explain the features
of quasiparticle interference in the superconducting 
cuprates~\cite{hanaguri}.

In this paper, we focused on a particular electron density 
corresponding to `hole doping' $x \approx 0.1$ on the
half filled state. However we  ignored
correlation effects that lead to the Mott state at half 
filling and also the possibility of competing phases
such as magnetic and charge order.
These effects are essential for any detailed understanding
of the underdoped cuprates. We touch upon this in our
discussion, next.

\section{Connection to experiments}

Having discussed our results we would like
to place them in the context of the pseudogap effect 
observed in the cuprates. 
There are two distinct regimes of `pseudogap physics'
as brought out by recent photoemission experiments. The
first pertains to the well studied underdoped regime,
where the proximity to the $x=0$ Mott insulator and
competing charge and spin order plays a role. The other,
more pertinent to us, occurs at relatively high doping,
$x \gtrsim 0.19$, where the `competing order' effects
are weak and one may be looking purely at $d$-wave
pairing fluctuations. Below, we first comment on the
underdoped regime, where we really cannot make any
quantitative comment, and then at the large doping
window.

\subsection{The underdoped regime}

The underdoped cuprates exhibit a wide variety of symmetry breaking 
phenomena including charge order, nematic order, breaking of time 
reversal as well as inversion symmetry, well above the superconducting 
transition temperature~\cite{chen}. The occurrence of these correlations 
also coincide with the presence of a dip in the DOS. 
There is no quasiparticle peak at the antinodal point above $T_c$ 
and a pseudogap appears on approaching $T_c$ marked by the existence 
of \textit{Fermi arcs}. The
length of Fermi arc continues to decrease and vanishes 
at $T_c$ when it gets transformed to a Fermi points. Our results show 
a qualitative agreement with this experimental feature though the size 
of gap decreases rapidly with temperature beyond $T_c$~\cite{kanigel2}.

Correlation effects in the underdoped
window can be approximately incorporated within
the Gutzwiller scheme that renormalize the hopping      
amplitude $t$ and pairing interaction $V$. The respective factors
are $g_t = 2x/(1 + x)$ and $g_s = 4/(1 + x)^2$, where $x$ is the
hole doping. The effective hopping vanishes as $x \rightarrow 0$
while the pairing interaction saturates. The ratio, 
$\tilde{V} /\tilde{t}$
= $(V /t) * 2/(x(1 + x))$. This suggests that the effective $V/t$ is
 enhanced significantly for $x = 0.1$ used in this work.

One can look at the consequence of this in two ways,
with similar qualitative conclusion. (i) Correlation effects would
suppress charge fluctuations, and the kinetic energy, in the doped  
Mott insulator. As suggested by pairing stiffness calculations in
the projected ground state~\cite{arun}, the $T_c$ would be   
strongly suppressed with respect to its ‘uncorrelated’ value, with
$T_c \rightarrow 0$ as $x \rightarrow 0$. This opens up a large
window between $T_c$ and the “pairing scale” decided by $V$.     
The pseudogap window increases because the $T_c$ gets     
lowered. Alternately, (ii) one can look at an ‘uncorrelated’ system
with $V/t$ now renormalised by $g_s/g_t$. It has been pointed    
out~\cite{mayr} that the temperature window for the pseudogap phase
increases with increasing $V/t$. Our Fig.1(b)     
shows the effect. Overall, the pseudogap window will widen     
at small $x$, compared to the estimate we make, due to     
the $T_c$ suppression caused by correlation effects. 

\begin{figure}[t]
\begin{center}
~~~
\includegraphics[scale = 0.72,angle = 0]{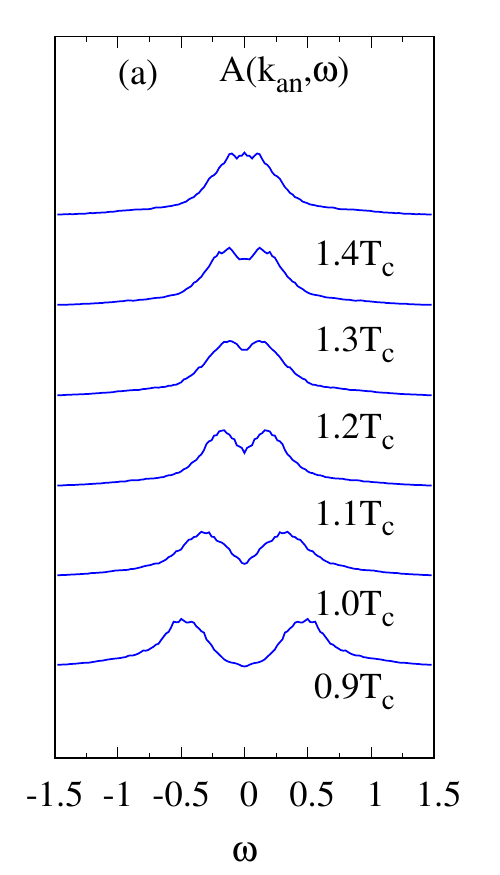}
\hspace{-0.4cm}
\includegraphics[scale = 0.22,angle = 0]{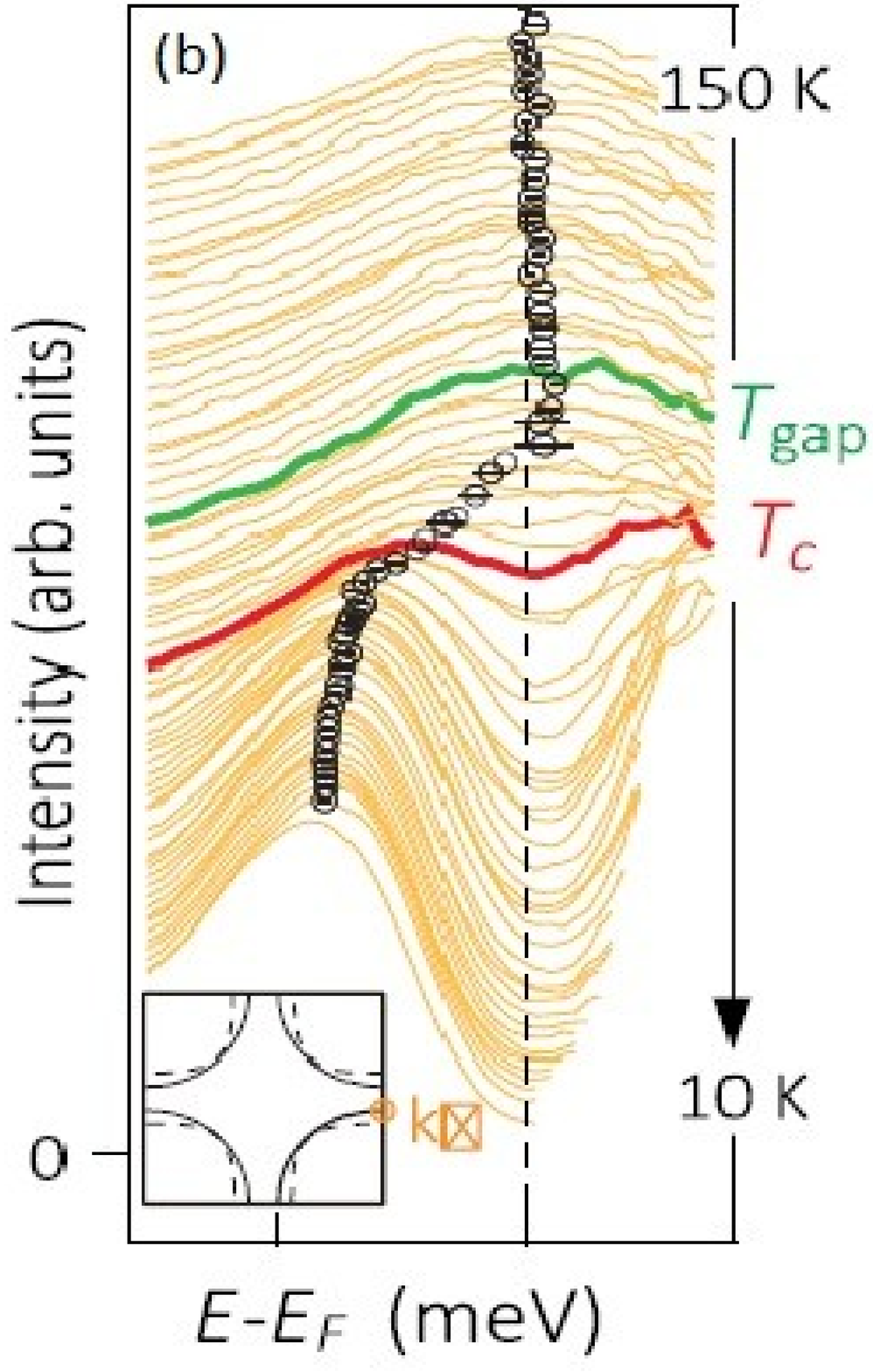}
\vspace{-5mm}
\end{center}
\caption{(a) Calculated and (b) experimentally determined antinodal 
quasparticle particle energy distribution curve $\mathcal{A}({\k}_{an}, 
\omega)$ as a function of temperature (Ref.[42]). 
Experimental data is for an overdoped sample.} \label{comp}
\end{figure}

\subsection{The large doping regime}

Beyond the underdoping window the state above
superconducting dome is expected to be a normal metal
marked by the presence of a well-defined quasiparticle
peak all along the Fermi surface. On the contrary, 
recent work~\cite{chen,he} indicates the persistence of
antinodal gap in the spectral function above $T_c$, a
feature suggested to be
intrinsic to a dwSC (Fig.\ref{comp} (b)). However, there
is a marked difference from the usual pseudogap phase, 
the lowering of temperature is accompanied first with a 
sharpening of quasiparticle peak before the gap appears 
close to but above the superconducting transition. 
Our calculations, which only takes $d$-wave 
correlations, and no competing order, into account shows a 
behaviour akin to what is observed in ARPES spectrum for
the antinodal point in the overdoped region (Fig.\ref{comp}(a)). 
First, we find
that the antinodal spectrum is gapped above $T_c$. 
Secondly, the gap closes rapidly with an increase in
temperature and disappears beyond $\sim 1.5T_c$ 
with the appearance of antinodal quasiparticles.

\section{Conclusions}

We have explored the possibility of a pseudogap phase
in a minimal microscopic model of $d$-wave superconductivity. 
We established the dependence of $T_c$ and the pseudogap
onset temperature $T^*$ on the pairing interaction $V$ and 
found that for $V$ typical of the cuprates the antinodal 
two-peak structure, with a shallow dip in between, 
persists in the 
momentum resolved spectrum upto $T^* \sim 1.5 T_c$. 
We observe that despite thermal fluctuations an essentially
nodal Fermi surface is seen for $T < T_c$, while for $T > T_c$
there is a Fermi arc feature, characterized by thermally 
broadened quasiparticle peaks, around the nodal points.
The arcs increase in length from $T_c$ to $T^*$, where 
they connect up to recover the normal state Fermi surface.
Around $T^*$ the two peak feature in the antinodal 
spectral function also collapses into a single peak
feature.  We provide a 
comprehensive map of the spectral function for
varying momentum and temperature.
The technical innovations used in this work can serve 
as the starting point for more elaborate models where the
role of competing channels, of density order or magnetism,
can be studied in conjunction with $d$-wave superconductivity.

\vspace{.3cm}

{\it Acknowledgement:}
D.K.S was supported through 
DST/NSM/R\&D$\_$HPC$\_$Applications/2021/14 
funded by DST of India and he would like to thank 
A. Akbari for useful discussions. Y. B. was supported
 through NRF Grant No. 2020-R1A2C2-007930 funded 
 by the National Research Foundation of Korea. 
 We acknowledge use of the HPC facility at HRI.

\end{document}